*GIZEM GULTEKIN-VÁRKONYI*[*]

## Why Avoid Generative Legal AI Systems? Hallucination, Overreliance, and their Impact on Explainability

This article argues that the deployment of generative AI systems in legal profession requires strong restraint due to the critical risks of hallucination and overreliance. Central to this analysis is the definition of Generative Legal AI (GLAI), an umbrella term for systems specifically adapted for the legal domain which is ranging from document drafting to decision support in criminal justice. Unlike traditional AI, GLAI models are built on architectures designed for statistical token prediction rather than legal reasoning, often leading to confabulations where the system prioritizes linguistic fluency over factual accuracy. These hallucinations obscure the reasoning process, while the persuasive, human-like nature of the output encourages professional overreliance. The paper situates these dynamics within the framework of European AI governance, arguing that the interaction between fabricated data and automation bias fundamentally weakens the principle of explainability. The article concludes that without effective mechanisms for meaningful human scrutiny, the routine adoption of GLAI poses significant challenges to judicial independence and the protection of fundamental rights.

### I. Introduction

The rapid integration of generative artificial intelligence (GenAI) into legal practice, as noted in many international reports specifically in the International Bar Association (IBA)[1] and the European Commission for the Efficiency of Justice (CEPEJ)[2], promises significant gains in efficiency but simultaneously introduces substantial epistemic and legal risks. Among the most salient problems are hallucinations, plausible yet fabricated outputs produced by large language models (LLMs), and overreliance, whereby legal professionals may defer uncritically to algorithmic recommendations perceived as authoritative. As systems such as ChatGPT increasingly permeate legal workflows, including legal research, drafting, and evidentiary analysis, these risks raise fundamental questions about explainability and accountability in legally consequential decision-making. Hallucinated legal authorities and confidently presented but inaccurate reasoning, as examples will be presented throughout the analysis in this paper, shows how generative systems can undermine the epistemic reliability required in judicial and legal contexts, particularly when users assume that AI-generated outputs are inherently trustworthy.

This paper examines the interaction between hallucinations and overreliance from the perspective of explainability obligations embedded in European AI governance. It first describes what are those GenAI tools that are specific to legal profession and how did they actually evolved within the historical lifecycle for AI systems. As identified in this paper as Generative Legal AI (GLAI) systems which is an umbrella term for those systems used specifically in legal domain for tasks starting from legal document drafting to decision support systems whih also might be used in criminal justice context. Then, the paper argues that hallucinations in GLAI systems obscure the reasoning processes underlying AI outputs, while overreliance reduces the likelihood that human actors critically interrogate those outputs generated by GLAI systems. Together, these dynamics weaken the practical realization of the explainability principles established in the General Data Protection Regulation (GDPR)[3] notably in relation to the Article 22 concerning automated decision-making. Further, the Artificial Intelligence Act (AI Act) [4]in, particularly the human oversight and transparency provisions specifically designated in the Articles 13 and 14 applicable to high-risk AI systems. By situating hallucinations and overreliance within the broader concept of explainability, the article analyses how cognitive biases, the persuasive fluency of generative models, and the opacity of large-scale machine learning systems interact to produce automation bias in legal decision environments. It ultimately argues that without effective mechanisms enabling meaningful scrutiny of AI-generated outputs, the regulatory guarantees of explainability and human oversight risk remaining largely formal, thereby posing challenges for judicial independence and the protection of fundamental rights in AI-lead legal practice.

---

[*] Senior Lecturer, University of Szeged, Faculty of Law and Political Sciences, International and Regional Studies Institute
[1] INTERNATIONAL BAR ASSOCIATION: *Guidelines and Regulations to Provide Insights on Public Policies to Ensure Artificial Intelligence's Beneficial Use as a Professional Tool*, September 2024. https://tinyurl.com/dv5rj4ks
[2] See; CEPEJ table prepared by the Resource Centre Cyberjustice and AI specific to identify AI applications in judicial systems: https://public.tableau.com/app/profile/cepej/vizzes

## II. *A brief history of connected AI, GenAI, and GLAI systems*

Rise of GenAI applications has a short but revolutionary history. Prior to early 2020, systems capable of generating human-like content in text, visuals, or audio were difficult, if not impossible, to conceive in real life application. Earlier tools executed predefined commands or provided fixed responses, differing markedly from current capabilities. This development actually did not occur suddenly, there are three major historical waves in AI history which shaped the development of this specific technology. The first vawe is considered in the academic literature as when was initiated by Alan Turing's 1950 paper "Computing Machinery and Intelligence"[5], which established AI foundations and introduced the Turing Test to evaluate machine-human response distinction. The period of the first vawe falls in between 1950s–1970s, amid post-World War II technological competition, introduced symbolic, rule-based AI system design. The ELIZA program developed by Joseph Weizenbaum for internal use at MIT[6] could be referred as one of the first prototypes of human-machine interactions through rule-based language processing. Some of the aims of the ELIZA system were basic keyword identification, discovery of minimal context, and output generation through responses which consist conversations.

The second wave which falls in between 1980s and early 2000s shifted to statistical machine learning for perception, prediction, and pattern recognition, constrained by hardware until advances in data, engineering, and military-industrial efforts. During this period, legal information retrieval systems were also rising with significant development phases. For example, the November-December 1966 issue of the American Bar Association's Jurometrics[7] magazine was dedicated to Legal Information Thru Electronic (a.k.a LITE) system. This system was developed in Colorado under the leadership of a group of military judges attached to the Air Force. In parallel with that, private database LexisNexis[8] was developed by American researchers (and still frequently used by legal professionals today), announced its online case law search feature in 1973. It is well known that legal information systems were also being developed in Europe at the time, for example the JURIS[9] database developed by the German Federal Ministry of Justice in the 1970s and privatized in the 1980s.

---

[3] Regulation (EU) 2016/679 of the European Parliament and of the Council of 27 April 2016 on the protection of natural persons with regard to the processing of personal data and on the free movement of such data, and repealing Directive 95/46/EC (General Data Protection Regulation), OJ L 119, 04/05/2016, p. 1–88.
[4] Commission Regulation (EU) 2024/1689 of the European Parliament and of the Council of 13 June 2024 laying down harmonised rules on artificial intelligence and amending Regulations (EC) No 300/2008, (EU) No 167/2013, (EU) No 168/2013, (EU) 2018/858, (EU) 2018/1139 and (EU) 2019/2144 and Directives 2014/90/EU, (EU) 2016/797 and (EU) 2020/1828 PE/24/2024/REV/1. OJ L, 2024/1689, 12.7.2024.
[5] TURING ALAN M.: *Computing machinery and intelligence*. Mind, 59(236), 1950, 433–460.
[6] WEIZENBAUM JOSEPH. *ELIZA—a computer program for the study of natural language communication between man and machine.* 1966, Commun. ACM 9, 1, 1966, 36–45. https://doi.org/10.1145/365153.365168

The third wave in the development of AI systems history falls in between early 2010s and 2020s which featured deep learning and foundation models. From 2020, the technology industry welcomed AI modelsthat were trained on extensive database information, information from internet and social media data for general tasks via objectives like next-token (letter to word prediction). This paved the way for the developers to train models with extensive data for adaptation without task-specific labels. Tools like ChatGPT transformed human-AI interaction. Legal AI systems, built on this framework, process legal data for knowledge discovery, predictive analytics, and related tasks. By 2025, 79% of law firms had integrated AI[10], even though full integration of AI systems in legal profession still in a slow phase.[11] Even though ChatGPT, Perplexity, and Claude function as GLAI due to their adaptability, offering benefits they pose significant challenges specifically to legal profession, as a result of their technical and practical settings. Before presenting these challenges which would explain why we do think that it is not recommended to utilize them in legal profession, we first would like highlight some of the technical setting which establishes the GenAI systems.

### 1. *Development of Generative AI Systems*

Since the advent of AI, theroetical systems which are capable of completing, if not surpassing human cognitive abilities has motivated researchers and industry leaders alike, including OpenAI CEO Sam Altman. This longstanding ambition, as extensively presented by Hao in her book[12], directly propelled OpenAI's development of ChatGPT, signifying a pivotal breakthrough amid prior stagnation in LLMs advancements despite mounting interest. Even though there are several other GPT tools available, OpenAI kicked the market off for development of different GenAI applications with this ambition.

The development of ChatGPT's architecture, particularly GPT-4, was further advanced through pretraining on an extensive corpus comprising hundreds of billions of words from books, articles, websites, and Wikipedia.[13] This corpus is estimated to be equivalent to the contents of 6,500 public libraries and exceeds one trillion parameters[14]. It is through this process that the model discerned statistical patterns in language, including subject-verb-object structures and connective words such as "and". Consequently, the model acquired the capacity for next-token prediction. Subsequent fine-tuning via supervised human interactions reinforced coherent, helpful responses, although challenges such as overreliance and hallucinated risks, explored in later discussions, tempered these gains.

---

[7] The editor of the journal at that time was ROBERT P. BIGELOW, who conducted extensive research on the use of computers in the field of law, authored the book "Computers and the Law: An Introductory Handbook" published in 1966, and served as president of the International Association for Technology Law from 1977 to 1979.
See: Jurimetrics Journal, 8, 2 (December 1966), pp. 83-85. https://www.jstor.org/stable/29761075?seq=1#metadata_info_tab_contents
[8] Azzolini lists Thomson Reuters, Reed Elsevier, and Wolters Kluwer as the top three global organizations offering tools to retreieve legal information. See; AZZOLINI JOHN: *The legal publishing world,* In Chandos Information Professional Series, Law Firm Librarianship, John Azzolini (ed.) Chandos Publishing, 2013, pp:165-189, https://doi.org/10.1016/B978-1-84334-708-8.50005-2.
[9] See; EXTER RITA, KAMMER MARTINA: (with updates by Sebastian Omlor in 2017) *"Legal Research in Germany between Print and Electronic Media - An Overview*
https://www.nyulawglobal.org/globalex/Germany1.html#VIII1
[10] KOCH MELISSA: *The AI Legal Landscape in 2025: Beyond the Hype*, 23 June 2025 https://www.akerman.com/en/perspectives/the-ai-legal-landscape-in-2025-beyond-the-hype.html.
[11] MICHELLE KIM: *AI might not be coming for lawyers' jobs anytime soon,* MIT Technology Review, 15 December 2025.
https://www.technologyreview.com/2025/12/15/1129181/ai-might-not-be-coming-for-lawyers-jobs-anytime-soon/
[12] HAO KAREN: *Empire of AI: Dreams and Nightmares in Sam Altman's OpenAI.* New York: Penguin Press, 2025. ISBN: 9780593657508.

ChatGPT currently boasts approximately 800 million weekly active users, representing a 5% market share.[15] This is not an isolated phenomenon; rather, it is indicative of broader generative AI ecosystems. These ecosystems encompass techniques for the creation of photorealistic images and videos (including deepfakes), diffusion models powering tools such as DALL-E, for data reconstruction, tools for 3D rendering, and discriminative models for predictive tasks such as recidivism assessment in criminal justice. The crux of OpenAI's innovation lies in the integration of these under a unified ChatGPT brand, thereby enabling multimodal generation across text, images, audio, and video. This multimodality together with its successfull performance metrics and results undoubtfully triggered many professional use cases, including GenAI use cases in legal profession.

## 2. *Generative AI Systems in Legal Profession: The GLAI existence*

In order to describe the current status of GLAI in real field applications, we refer to two international reports prepared by international organizations relevant to legal profession. International reports by the CEPEJ[16] and the IBA[17] offer two complementary perspectives on the current deployment of AI in the legal domain. Although neither report is specifically designed to classify generative AI technologies, a closer examination of their categories reveals several functional areas in which generative AI systems have become particularly visible. These areas explicity point the broader trend whereby LLM based systems are integrated into existing legal technology infrastructures rather than emerging as entirely separate categories:[18]

> (1) One of the most prominent domains in which generative AI appears is legal information discovery and document analysis. The CEPEJ framework refers to this category as "document search, review and large-scale discovery,"[19] while the IBA describes a comparable function as "legal information discovery."[20] Traditionally, such systems were primarily designed to retrieve and structure legal information from large databases of legislation and case law. However, recent GenAI systems increasingly extend these capabilities by producing summaries of judgments, extracting key legal arguments, or generating structured explanations of legal texts.[21] Systems based on LLMs, such as ChatGPT or Microsoft Copilot, are frequently used in this context to synthesize legal materials and assist professionals in navigating complex legal corpora. These developments align with broader research on AI-assisted legal analytics and natural language processing in law.[22]

---

[13] O'HARA MATTHEW J.: *"I, for One, Welcome Our New" AI Jurors: ChatGPT and the Future of the Jury System in American Law,* International Journal of Law, Ethics, and Technology, 4 (4), 2025, pp. 50-84. https://ijlet.org/wp-content/uploads/2025/01/IJLET-4.3.2.pdf
[14] BASTIAN MATTHIAS: *GPT-4 has more than a trillion parameters – Report,* The Decoder, 25 March 2023. https://the-decoder.com/gpt-4-has-a-trillion-parameters/
[15] BELLAN REBECCA: *Sam Altman says ChatGPT has hit 800M weekly active users.* TechCrunch, 6 October 2025. https://techcrunch.com/2025/10/06/sam-altman-says-chatgpt-has-hit-800m-weekly-active-users/
[16] CEPEJ table, n.d.
[17] IBA, 2024
[18] TERZIDOU KALLIOPI: *Generative AI Systems in Legal Practice Offering Quality Legal Services While Upholding Legal Ethics.* 2025, International Journal of Law in Context 21, no. 3, 2025, pp.431–52. https://doi.org/10.1017/S1744552325000047.
[19] CEPEJ table, n.d.
[20] IBA, 2024

(2) A second domain where GenAI tools are visible is legal drafting and document generation. While this category appears primarily in the IBA report, its functional scope overlaps with several CEPEJ decision-support tools that assist judges and court staff in summarising case materials or generating draft reasoning. GenAI systems are particularly suited to drafting tasks because of their ability to produce structured text outputs, including pleadings, contracts, or legal correspondence.[23] In practice, many contemporary legal technology platforms integrate drafting capabilities with research functions, enabling users to generate documents based on retrieved legal authorities or previously analysed case materials.[24]

(3) GenAI is also increasingly present in decision-support and litigation analytics tools. The CEPEJ framework identifies systems aimed at "decision support" and "prediction of litigation outcomes,"[25] while the IBA report describes similar technologies under the category of "outcome estimation."[26] Historically, these tools relied on statistical modelling of judicial decisions. More recently, GenAI systems have been incorporated to interpret case data, produce narrative explanations of analytical results, and summarise patterns detected within judicial datasets. Although such systems do not (intend to) replace judicial decision-making, they may inform legal reasoning by providing structured insights into precedent patterns or litigation trends.

(4) Another category in which GenAI plays a visible role is public and professional legal information and assistance services. Both reports describe systems designed to provide legal guidance or procedural information to non-expert users, usually via chatbots. The recent emergence of conversational GenAI tools, such as Grok and Mistral AI, has significantly expanded in this domain. These systems can respond to natural language queries, explain legal procedures, and simplify complex legal documents for broader audiences. Their accessibility has contributed to the rapid diffusion of GenAI within the justice sector, particularly in contexts related to access to justice and legal information. These tools can

---

[21] MICROSOFT: *Using Copilot in Legal*. https://adoption.microsoft.com/en-us/scenario-library/legal/
[22] SOURDIN TANIA: *Judges, Technology and Artificial Intelligence*, Elgar Law, Technology and Society, 2021, e-book, pp. 275-276.
[23] REGALIA, JOSEPH: *From Briefs to Bytes: How Generative AI is Transforming Legal Writing and Practice*. Scholarly Works, 1457, 2024. https://scholars.law.unlv.edu/facpub/1457
[24] See for further functions: Thomson Reuters CoCounsel Legal: https://legal.thomsonreuters.com/en/products/cocounsel-legal and a commercial tool developed for same purposes: Harvey AI: https://www.harvey.ai/
[25] OBERTO GIACOMO: *Artificial Intelligence And Judicial Activities: The Position Of The European Commission For The Efficiency Of Justice (CEPEJ)*. In: Artificial Intelligence: its Impact on the Judicial Activities" conference organised by tthe International Association of Judges (IAJ) in co-operation with the Judicial Officers Association of South Africa (JOASA). Cape Town, Sunday 20th October, 2024. https://tinyurl.com/2s367zfy
[26] The question whether AI tools could serve for legal prediction or what they actually do falls under one of the either categories under the court settings which is relevant to judgment categorization, outcome forecasting or outcome identification have been discussed extensively here:
MEDVEDEVA MASHA, WIELING MARTIJN, VOLS MICHEL: *Rethinking the field of automatic prediction of court decisions*. Artif Intell Law, 31, pp.195–212, 2023. https://doi.org/10.1007/s10506-021-09306-3

      target a specific field such as tax law consultation or child arrangement disputes[27] to general legal consultation bots claiming to be expert in all jurisdictions in the world.[28]

(5) Finally, GenAI capabilities also appear indirectly in online dispute resolution (ODR) systems. While ODR platforms predate generative AI, newer implementations increasingly integrate conversational interfaces and automated drafting support to assist parties in preparing claims, responses, or settlement proposals. In this sense, GenAI functions as an enabling layer within digital dispute resolution infrastructures rather than as a standalone dispute resolution mechanism. [29]

Taken together, the comparative approach we took to the CEPEJ categorization and IBA reports suggests that GenAI systems are visible in at least five functional domains: (1) legal information discovery, (2) legal drafting and document generation, (3) decision-support and litigation analytics, and (4) public-facing legal assistance tools, and the least GLAI affected category, (5) ODR. Rather than forming a separate category of legal technology, GenAI increasingly operates as a multiple domain capability embedded within existing AI systems for general purposes, but still, reshaping how legal information is processed, generated, and communicated within the digital justice lifecycle.

## III. The Problem of Confabulation

„Just feed it a prompt like "write me a securities opinion with lots of citations about scienter" and see what happens."[30]

In late July 2025, a striking incident unfolded in the U.S. District Court for New Jersey when Judge Julien Xavier Neals withdrew an entire opinion after it was found to contain visible factual errors deriving from fabricated quotes, mischaracterized rulings, references to a case that seemingly never existed, and statements falsely attributed to defendants. The judge, within the scope of his professionalism, could detect them and took measures to avoid these illusionary cases to enter in the U.S. case law database.

The phenomenon of confabulation, frequently referred as hallucination, represents a critical epistemic failure in the deployment of LLMs within the legal and other scientific domains. Unlike human cognitive errors, these outputs are systematic fabrications necessitated by the underlying architecture of generative AI, which prioritizes statistical token frequency over empirical "ground truth"[31]. Emsley (2023) states that the term hallucination is an anthropomorphic misnomer, rather, these models produce stochastic fabrications because they are designed for linguistic fluency rather than factual accuracy.[32] There is a simple reason why behind LLMs hallucinate and it requires mostly a technical explanation. As stated by Mittelstadt et al.,[33] these models are trained on massive, unstructured datasets scraped from the internet, the training corpora are inherently resistance to truth and containing a mixture of empirical facts, creative fiction, and common bias features. Once they met LLMs probabilistic mix-matches in opaque data processing ststems, it results in what has been termed as careless speech defined as a form of communication that lacks a duty to the truth and presents subjective or non-existent data as objective fact.[34]

---

[27] See, Justice Chatbots provided by the UK government with the aim of supporting citizens legal questions and requests: https://ai.justice.gov.uk/ai-in-action/justice-chatbots
[28] The following tool might be embedded in the firm's website to collect training data as well: https://lawconnect.com/chat
[29] KATSH ETHAN, RABINOVICH-EINY ORNA: *Digital Justice: Technology and the Internet of Disputes*. Oxford University Press, 2017, pp. 33–34. DOI:10.1093/acprof:oso/9780190464585.001.0001
[30] SCARCELLA MIKE: *Two US judges withdraw rulings after attorneys question accuracy*. 29 July 2025 Reuters. https://www.reuters.com/legal/government/two-us-judges-withdraw-rulings-after-attorneys-question-accuracy-2025-07-29/
[31] WACHTER SANDRA, MITTELSTADT BRENT, RUSSELL CHRIS: *Do large language models have a legal duty to tell the truth?*. R Soc Open Sci. 11 (8): 240197, 2024. https://doi.org/10.1098/rsos.240197

This problem is highly relevant to explainability principle which is widely discussed in the legal literature aiming to explain the principle embedded in the EU AI governance. For several years, legal literature hosted several discussions on the possibility of exaplainable AI systems and if this is possible, then how these explanations should be like. In the literature, algorithmic explainability has usually been related to either explanaiton of the processes what the AI model goes under[35] or explaining the outcome which the AI model reached after its performance.[36] Specific to GLAI systems, such as the ones being used in criminal justice brought a new perspective to these discussions. For example, Ersöz et al.'s (2025)[37] research identfiy the fact that even though there exist several methods to enhance explanaibility in crime prediction models, they are still lack of necessary ethical settings but also including fairness, accountability, explainability, interpretability, security, and privacy. Even under circumstances when the data is real and not fabricated, the problem with explainability cannot be overcome, the situation with the fabriacted data would reinforce this problem by feeding back the non-existed data to the model for future references.

The explainability challenge, besides its technical settings, intensifies under the GDPR and AI Act, both of which mandate transparency and explainability for AI systems. Article 22 of the GDPR grants individuals the right not to be subject to automated decisions with legal effects (or similarly significant effects) unless explicit safeguards apply, while Recital 71 emphasizes the need for "meaningful information about the logic involved" in such decisions. The AI Act further classifies certain legal AI applications (e.g., crime prediction) as high-risk, requiring detailed technical documentation, risk assessments, and human oversight to ensure outputs are traceable and comprehensible (Article 13). These provisions demand robust explainability to enable affected parties to understand, contest, and seek redress for AI-driven decisions. Hallucinations erode explainability by introducing unverifiable data into decision pipelines, rendering process tracing unreliable. As a result, a decision maker cannot credibly justify outcomes derived from nonexistent facts, even if real data proves challenging. Even with authentic datasets, explainability challenge remains since fabricated inputs exacerbate this, as models may perpetuate errors in future iterations, violating GDPR's accuracy principle (Article 5(1)(d)) and AI Act obligations for data governance (Article 10).[38]

---

[32] EMSLEY ROBIN: *ChatGPT: these are not hallucinations–they're fabrications and falsifications.* Schizophrenia, 2023, 9:52. doi: 10.1038/s41537-023-00379-4.
[33] MITTELSTADT BRENT, WACHTER, SANDRA, RUSSELL, CHRIS: *To protect science, we must use LLMs as zero-shot translators.* Nat Hum Behav, 7, 2023,1830–1832 https://doi.org/10.1038/s41562-023-01744-0
[34] WACHTER, MITTELSTADT, RUSSELL.
[35] GÓRSKI LUKASZ, RAMAKRISHNA SHASHISHEKAR: *Explainable artificial intelligence, lawyer's perspective.* In: Proceedings of the Eighteenth International Conference on Artificial Intelligence and Law (ICAIL '21). Association for Computing Machinery, New York, NY, USA, 2021, 60–68. https://doi.org/10.1145/3462757.3466145
[36] MEDVEDEVA MASHA: *Law is ready for AI, but is AI ready for law?* Internet Policy Review, 24 July 2025, https://policyreview.info/articles/news/ai-ready-law/2023
[37] ERSÖZ FILIZ, ERSÖZ TANER, MARCELLONI FRANCESCO, ET AL.: *Artificial Intelligence in Crime Prediction: A Survey With a Focus on Explainability.* In: IEEE Access,13, pp. 59646-59674, 2025, doi: 10.1109/ACCESS.2025.3553934.

Specific to the scope of GLAI practices, it is important to note some instances where hallucinations infringe upon individual rights, such as the right to privacy, ultimately resulting in legal action operating the relevant legislation. A new complaint filed by NOYB against OpenAI claims that ChatGPT allegedly produced a defamatory hallucination falsely claiming a Norwegian individual had murdered his children.[39] Subsequent technical examinations found no evidence to support the claim, supporting the assumption in a way that the output was entirely fabricated despite incorporating accurate personal details. As these errors become more common, they create a major problem for experts in fields like law, journalism, and medicine. Verifying every piece of information takes a lot of time and resources. If professionals rely too much on AI without careful checking, the quality of their work will likely drop. In some cases, the effort needed to check the AI's work might be even greater than doing the work manually from the start.[40]

From a legal AI perspective, such an incident prooves a built-in high risk where GenAI errors mislead professionals into making important decisions based on false information. Since the AI Act has not yet entered into forced entirely, and there is no case analyzed yet under its rules, therefore we cannot present a concrete evidence under the AI Act's porivision. Connected to the hallucinations, let us imagine that there was a law enforcement official believing in the Norwegian applicant's case and takes further actions in line with it. Here another risk arises, which refers to the fact that AI systems presenting these errors often in a very confident and professional tone. This makes it difficult for human users to stay skeptical or notice that the facts are wrong. Erosion of professional critical thinking is the other reason why GLAI might not be considered as a trustful companion.

## IV.     The Problem of Overreliance

The growing dependence on AI systems in different areas of professional[41] and private life[42] has its root to some of marketing strategies and the remarkable capacity of GenAI tools to mimic human speech and behavior. These systems are frequently portrayed as authoritative sources of insight, reinforcing a phenomenon known as overreliance, wherein users attribute unwarranted credibility to algorithmic outputs, often mistaking them for infallible truth.[43] By incorporating humanoid characteristics, such as expressive language and responsive interactions, GenAI blurs the distinction between simulation and genuine cognition, leading individuals to overlook the underlying computational mechanisms. This predisposition is particularly pronounced among legal professionals, including judges and lawyers, who may perceive AI as inherently superior due to its efficiency, impartiality, and speed attributes highlighted in prior discussions.[44] However, such perceptions are misguided, AI neither surpasses human judgment in all contexts nor warrants uncritical acceptance, especially given its propensity for hallucinations. Plausible yet fabricated content that users readily accept as factual, is one of the biggest threats that may undermine rational decision-making which is a principle of (legal) professionalizm.

---

[38] CHRISTAKIS THEODORE: *AI Hallucinations and Data Subject Rights under the GDPR: Regulatory Perspectives and Industry Responses*, 2024. https://hal.univ-grenoble-alpes.fr/hal-04844898v1

[39] NOYB: *AI hallucinations: ChatGPT created a fake child murderer*. 20 March 2025. https://noyb.eu/en/ai-hallucinations-chatgpt-created-fake-child-murderer

[40] DELL'ACQUA FABRIZIO, MCFOWLAND EDWARD, MOLLICK ETHAN, ET AL: *Navigating the Jagged Technological Frontier: Field Experimental Evidence of the Effects of AI on Knowledge Worker Productivity and Quality*. Harvard Business School Working Paper, No. 24-013, September 2023.

[41] QIANQIAN WU: *How AI Dependence influences employees' Time Theft?* In: Proceedings of the 2025 2nd International Conference on Digital Society and Artificial Intelligence (DSAI '25). Association for Computing Machinery, New York, NY, USA, 2025, 96–102. https://doi.org/10.1145/3748825.3748542

[42] GERLICH MICHAEL: *AI Tools in Society: Impacts on Cognitive Offloading and the Future of Critical Thinking*. Societies 15, 1 (6), 2025. https://doi.org/10.3390/soc15010006

[43] PEARSON JOE, DROR ITIEL E., JAYES EMMA, *et al. Examining human reliance on artificial intelligence in decision making*. Sci Rep, 16, 2026, 5345. https://doi.org/10.1038/s41598-026-34983-y

This overreliance can be traced to well-established psychological and social paradigms, providing a coherent framework for understanding its emergence. Masahiro Mori's concept of the "uncanny valley" which is dated back to 1970 explains how increasing human likeness in artificial entities elicits familiarity up to a point, beyond which revulsion arises, but critically, high fidelity initially deceives users into anthropomorphic attribution. Building on the grounds of this foundation, Reeves and Moon's „Computers as Social Actors" paradigm[45] claims that humans instinctively respond to machines mimicking social cues, attributing human-like qualities through automatic, unreflective processes despite the absence of biological sentience. La Rosa and Danks[46] extend this lineage, arguing that sustained trust arises when AI fulfills perceived roles effectively, reinforcing reliance. All these researches and claims show us that actually, persuasive abilities of GenAI systems could have been avoided if the systems were designed in line with these evidences. However, ChatGPT expresses simulated emotions, issues apologies for errors upon wrong answers based on user feedbacks, and solicits feedback via like mechanisms that double as reinforcement learning inputs based on emotional ties. Such interactions have appeared in severe real world consequences such as the dangers of conflating output with reality, besides the ones we presented about the lawyers presenting hallucinated cases. In April 2025, a U.S. high school student died by suicide following prolonged conversations with ChatGPT, prompting lawsuits against OpenAI, which deflected responsibility to user misuse.[47]

In legal contexts, overreliance amplifies procedural and systemic risks, with empirical examples highlighting the erosion of professional oversight. Legal filings incorporating ChatGPT generated hallucinations, as we provided before, have surfaced across jurisdictions outside of the U.S. A recent Turkish case involved an enforcement office prioritizing the model's rent calculation over undisputed documentary evidence, bypassing critical verification.[48] Habitual delegation of routine tasks atrophies core competencies like legal reasoning, positioning lawyers as essential checks alongside developers[49]. Ghasemaghaei and Kordzadeh (2024)

---

[44] GROSSMAN MAURA R, GRIMM PAUL W, BROWN DANIEL G, ET AL.:*The GPTJudge: Justice in a Generative AI World*. Duke Law & Technology Review, 23, 2023 1 (34).

[45] NASS CLIFF, MOON YOUNGME: *Machines and Mindlessness: Social Responses to Computers,* Journal of Social Issues, 56, 2000, pp.81-103.

[46] LAROSA EMILY, DANKS DAVID: *Impacts on Trust of Healthcare AI*. In: Proceedings of the 2018 AAAI/ACM Conference on AI, Ethics, and Society, New York, NY, USA: ACM (AIES '18), pp. 210–215. doi: 10.1145/3278721.3278771

[47] BOOTH ROBERT : *ChatGPT firm blames boy's suicide on 'misuse' of its technology*, The Guradian, 26 November 2025. https://www.theguardian.com/technology/2025/nov/26/chatgpt-openai-blame-technology-misuse-california-boy-suicide
Family of the deceased boy filed law suits in several jurisdictions ans the cases are ongoing: Matthew Raine et al vs. OpenAI, Inc. https://www.documentcloud.org/documents/26075676-raine-v-openai

[48] Thanks to Melis Aksoy for sharing this incident on LinkedIn: https://www.linkedin.com/feed/update/urn:li:activity:7317896695088865280

[49] GRAVETT WILLEM: *Is the Dawn of the Robot Lawyer upon Us? The Fourth Industrial Revolution and the Future of Lawyers.* Potchefstroom Electronic Law Journal, 23, 2020, pp. 1-37. https://perjournal.co.za/article/view/6794/10195

[50]offer the "obedience to authority" framework, describing AI as an authoritative entity whose recommendations users follow with diminished guilt, even when discriminatory. Their study of 122 U.S. managers demonstrated compliance with algorithmic biases in hiring, attributing this to outsourced moral responsibility which we could call as snowball effect, where initial deference begets habitual reliance. Criminal justice tools like the one widely known COMPAS tool has been in the U.S. could further support these claims. Designed to score recidivism risk (1-10) rapidly, it swayed sentencing in State v. Loomis (2016)[51], where the defendant challenged opaque data, gender proxies, and lack of individualization. Courts upheld its use for some time, absent due process violation but cautioned against determinism, dissenting opinions decried "black box" opacity shielded by trade secrets, potentially pressuring judicial independence. Notably, COMPAS evaluates 137 variables (e.g., priors, substance use)[52], yet comparable accuracy derives from just age and convictions.[53] Decision-makers, being misguided by efficiency promises, trust these scores implicitly, allowing hallucinations or biases to infiltrate human judgments without contest.

Article 14(3) of the EU AI Act directly addresses overreliance risks in legal AI systems by mandating specific human oversight safeguards for high-risk applications, such as crime prediction and recidivism assessment tools classified under Annex III, high-risk AI systems. The provision requires deployers to ensure operators "remain aware of the possible tendency of automatically relying or over-relying on the output produced by a high-risk AI system (automation bias)," particularly where systems provide "information or recommendations for decisions to be taken by natural persons." However, and once again, since AI Act has not yet fully entered into force, and there is no guarante for its full applicability is any soon, no further concrete interpretation is possible at the moment.

## V.    A possible solution: Reinforcing exaplainations, but how?

Explanations may emerge as a promising mitigation strategy for both AI hallucinations and overreliance, particularly in legal systems, by restoring transparency to otherwise opaque generative processes, when combined with the existing legislation which legall bound the providers and deployers of such systems. "Explainable by design" concept which we offer as a solution here must be understood as complementary to other principles referred in law, in particular in the AI Act. For example, while explainability enhancing techniques help stakeholders understand and interpret decision logic and causal relationships in AI output, robustness principle (Article 15) ensures performance remains reliable under variable inputs and environments.[54]  Further, Papagiannidis et al. (2025)[55] position explainability and transparency as central pillars of responsible AI governance, integral to fairness and accountability. Akbarighatar (2025)[56] identifies five interrelated capability themes aligned with explainability enhancement: understandable models, bias remediation, responsiveness to inquiries, harmlessness through ethical design, and orientation toward the common good.

---

[50] GHASEMAGHAEI MARYAM, KORDZADEH NIMA: *Understanding how algorithmic injustice leads to making discriminatory decisions: An obedience to authority perspective*. Information & Management, 61 (2), 2024, 103921, ISSN 0378-7206.
https://doi.org/10.1016/j.im.2024.103921.
[51] 881 N.W.2d 749 (Wis. 2016)
[52] BRENNAN TIM, DIETRICH WILLIAM, EHRET BEATE: *Evaluating the Predictive Validity of the Compas Risk and Needs Assessment System*. Criminal Justice and Behavior, 2020, pp. 21–40.
[53] DRESSEL JULIA, FARID HANY: *The accuracy, fairness, and limits of predicting recidivism*. Sci. Adv. 4, 2018, 1–5. DOI: 10.1126/sciadv.aao5580.
[54] CHANDER BHANU, JOHN CHINJU, WARRIER LEKHA, ET AL.: *Toward Trustworthy Artificial Intelligence (TAI) in the Context of Explainability and Robustness*. ACM Comput. Surv. 57, 6, Article 144 (June 2025), p. 49. https://doi.org/10.1145/3675392
[55] PAPAGIANNIDIS EMMANOUIL, MIKALEF PATRICK, CONBOY KIERAN: *Responsible artificial intelligence governance: A review and research framework*. The Journal of Strategic Information Systems, 2025, 34 (2),101885. https://doi.org/10.1016/j.jsis.2024.101885.
The authors also point the fragmented literature on ethical AI design and operationalization, and the limited clarity on how responsible AI norms can be operationalized during model design, deployment, monitoring, and evaluation. This points clearly that is not that easy and straightforward, but again, there are efforts in this direction.

In their work, Vasconcelos et al.[57] contest prior researches as they claim that explanations for AI predictions does not reduce such overreliance, and the authors aim to identify conditions under which explanations can actually mitigate this problem. The authors develop a cost–benefit framework, arguing that users strategically decide whether to engage with an AI's explanation based on the cognitive costs of verification and the benefits of relying on the AI tool. According to the result of their research, explanations can reduce overreliance when they sufficiently lower the effort required to verify the AI's output or when incentives motivate users to engage more deeply with the task. Nevertheless, standardized criteria for evaluating explainability in legal AI remain underdeveloped.[58] To overcome with this, explainability must be, first of all, domain-aware. Building on these outputs, we can suggest that explanations require adaptation for legal reasoning, as legal decision-making demands contestable, defeasible, and normative justifications rather than generic feature importance scores. Ordinary explainability techniques may fail in legal contexts because they are not specifically designed to capture law's argumentative structure, where decisions must withstand adversarial scrutiny and align with statutory interpretation principles. Our final suggestion, if is not to be perceived a drastic one, to rather not to use these tools in legal contexts unless the problem of hallucination and overreliance are completely solved which is impossible from our point of view.

*VI. Conclusion*

The evolution of AI into the era of LLMs and GenAI has impacted uses cases in legal profession which we call in this article as GLAI. These systems, while efficient, operate in a black box setting that detect statistical regularities rather than legal relevance. This technical reality creates a dangerous synergy between hallucinations and overreliance that directly undermines the core legal principle of explainability. Hallucinations erode explainability by introducing unverifiable, non-existent data into decision pipelines, making it impossible for a legal professional to credibly justify an outcome. Simultaneously, the anthropomorphic and persuasive fluency of GLAI triggers a social overreliance, leading legal professionals to outsource their moral and professional responsibility to an algorithm. This dual failure renders regulatory safeguards, such as those found in the GDPR and the AI Act, largely formal rather than substantive. While "explainable by design" strategies may offer a path forward, they must be made domain-aware to account for the unique argumentative structure of law. Until the risks of confabulation and overreliance are fully resolved which appears improbable at the current

---

[56] AKBARIGHATAR, POURIA: *Operationalizing responsible AI principles through responsible AI capabilities*. AI Ethics, 5, pp.1787–1801 2025. https://doi.org/10.1007/s43681-024-00524-4

[57] VASCONCELOS HELENA, JÖRKE MATTHEW, GRUNDE-MCLAUGHLIN MADELEINE, ET AL: *Explanations Can Reduce Overreliance on AI Systems During Decision-Making*. In: Proc. ACM Hum.-Comput. Interact. 7, CSCW1, Article 129 (April 2023), p.38 https://doi.org/10.1145/3579605

[58] MANSI GENNIE, KARUSALA NAVEENA, RIEDL MARK: *Legally-Informed Explainable AI*. In: Philipp Wintersberger (ed.), Proceedings of the 5th HCXAI Workshop @ CHI (Yokohama, Japan, April 26-May 1, 2025)

stage, the routine use of GLAI in legally consequential decision-making remains incompatible with the legal and ethical standards of the legal profession.

*GULTEKIN-VÁRKONYI GIZEM*[*]

**Miért kerüljük a generatív jogi mesterségesintelligencia-rendszereket? A hallucináció, a túlzott hagyatkozás és ezek magyarázhatóságra gyakorolt hatása**

E tanulmány amellett érvel, hogy a generatív mesterségesintelligencia-rendszerek (MI) jogi szakmában történő alkalmazása komoly önmérsékletet igényel a hallucináció és a túlzott hagyatkozás kritikus kockázatai miatt. Az elemzés központi eleme a Generatív Jogi MI (Generative Legal AI – GLAI) meghatározása, amely gyűjtőfogalomként a kifejezetten jogi területre adaptált rendszereket jelöli – a dokumentumszerkesztéstől kezdve egészen a büntető igazságszolgáltatási döntéstámogatásig. A hagyományos MI-vel ellentétben a GLAI-modellek olyan architektúrákra épülnek, amelyeket statisztikai token-predikcióra, nem pedig jogi érvelésre terveztek; ez gyakran konfabulációkhoz vezet, ahol a rendszer a nyelvi gördülékenységet részesíti előnyben a ténybeli pontossággal szemben. Ezek a hallucinációk elfedik az érvelési folyamatot, miközben a kimenet meggyőző, emberi jellegű stílusa szakmai túlzott hagyatkozásra ösztönöz. A cikk az európai MI-irányítás keretrendszerébe ágyazza ezeket a dinamikákat, hangsúlyozva, hogy a fiktív adatok és az automatizációs elfogultság (automation bias) interakciója alapjaiban gyengíti a magyarázhatóság elvét. A tanulmány végkövetkeztetése szerint az érdemi emberi felügyelet hatékony mechanizmusai nélkül a GLAI rutinszerű alkalmazása jelentős kihívások elé állítja a bírói függetlenséget és az alapvető jogok védelmét.

---

[*] adjunktus, SZTE ÁJTK NRTI